\documentclass[12pt]{iopart}
\include{epsf}
\usepackage{amssymb}

\newcommand{\cut}[1]{{}}

\def\vy{\mbox{\boldmath{$y$}}}
\def\vb{\mbox{\boldmath{$b$}}}
\def\vtau{\mbox{\boldmath{$\tau$}}}
\def\vsigma{\mbox{\boldmath{$\sigma$}}}
\def\vs{\mbox{\boldmath{$s$}}}
\def\vomega{\mbox{\boldmath{$\omega$}}}
\def\vxi{\mbox{\boldmath{$\xi$}}}
\def\tL{{\tilde L}}
\def\tC{{\tilde C}}

\newcommand{\sign}{\mathrm{sign}}
\newcommand{\erfc}{\mathrm{erfc}}

\renewcommand{\setminus}{\smallsetminus}

\begin{document}

\title{Sparsely-spread CDMA - a statistical mechanics based analysis}

\author{Jack Raymond and David Saad}

\address{
Neural Computation Research Group, Aston University, Aston Triangle, Birmingham, B4  7EJ
}
\ead{jack.raymond@physics.org}
\begin{abstract}

Sparse Code Division Multiple Access (CDMA), a variation on the
standard CDMA method in which the spreading (signature) matrix
contains only a relatively small number of non-zero elements, is
presented and analysed using methods of statistical physics. The
analysis provides results on the performance of maximum likelihood
decoding for sparse spreading codes in the large system limit. We
present results for both cases of regular and irregular spreading
matrices for the binary additive white Gaussian noise channel
(BIAWGN) with a comparison to the canonical (dense) random
spreading code.
\end{abstract}
\pacs{64.60.Cn, 75.10.Nr, 84.40.Ua, 89.70.+c}
\ams{68P30,82B44,94A12,94A14}
\date{\today}

\maketitle

%%%%%%%%%%%%%%%%%%%%%%%%%%%%%%%%%%%%%%%%%%%%%%%%%%%%%%%
\section{Background}
%%%%%%%%%%%%%%%%%%%%%%%%%%%%%%%%%%%%%%%%%%%%%%%%%%%%%%%

The area of multiuser communications is one of great interest from
both theoretical and engineering perspectives~\cite{Verdu:MD}.
Code Division Multiple Access (CDMA) is a particular method for
allowing multiple users to access channel resources in an
efficient and robust manner, and plays an important role in the
current preferred standards for allocating channel resources in
wireless communications. CDMA utilises channel resources highly
efficiently by allowing many users to transmit on much of the
bandwidth simultaneously, each transmission being encoded with a
user specific signature code. Disentangling the information in the
channel is possible by using the properties of these codes and
much of the focus in CDMA research is on developing efficient
codes and decoding methods.

In this paper we study a variant of the original method, sparse CDMA, where the spreading matrix contains only a relatively small number of non-zero elements as was originally studied and motivated in~\cite{Yoshida:ASS}. While the straightforward application of sparse CDMA techniques to uplink
multiple access communication is rather limited, as it is difficult to synchronise the sparse transmissions from the various users, the method can be highly useful for frequency and time hopping. In frequency-hopping code division multiple access (FH-CDMA), one repeatedly switches frequencies during radio transmission, often to minimize the effectiveness of interception or jamming of telecommunications. At any given time step, each user occupies a small (finite) number of the (infinite) $M$-ary frequency-shift-keying (MFSK) chip/carrier pairs (with gain $G$, the total number of chip-frequency pairs is $MG$.) Hops between available frequencies can be either random or preplanned and take place after the transmission of data on a narrow frequency band. In time-hopping (TH-)CDMA, a pseudo-noise sequence defines the transmission moment for the various users, which can be viewed as sparse CDMA when used in an ultra-wideband impulse communication system. In this case the sparse time-hopping sequences reduces collisions between transmissions.

This study follows the seminal paper of Tanaka~\cite{Tanaka:SMA},
and other recent extensions~\cite{Guo:MDSP}, in utilising the
replica analysis for randomly spread CDMA with discrete inputs,
which established many of the properties of random densely-spread
CDMA with respect to several different detectors including Maximum
A Posteriori (MAP), Marginal Posterior Maximiser (MPM)
and minimum mean square-error (MMSE). Sparsely-spread CDMA differs
from the conventional CDMA, based on dense spreading sequences, in
that any user only transmits to a small number of chips (by
comparison to transmission on all chips in the case of dense
CDMA). The sparse nature of this model facilitates the use of
methods from statistical physics of dilute disordered
systems~\cite{Nishimori:SP,Mezard:SGT} for studying the properties
of typical cases.

The feasibility of sparse CDMA for transmitting information was recently demonstrated~\cite{Yoshida:ASS} for the case of real (Gaussian distributed) input symbols by employing a Gaussian effective medium approximation; several results have been reported for the case of random transmission patterns. In a separate recent study, based on the belief propagation inference algorithm and a binary input prior distribution, sparse CDMA has also been considered as a route to proving results in the densely spread CDMA~\cite{Montanari:ABP}. In addition, this study demonstrated the existence of a {\it waterfall} phenomenon comparable to the dense code for a subset of ensembles. The waterfall phenomenon is observed in decoding techniques, where there is a dynamical transition between two statistically distinct solutions as the noise parameter is varied. Finally we note a number of pertinent studies concerning the effectiveness of belief propagation as an MPM decoding method~\cite{Kabashima:SMA,Neirotti:IMP,Montanari:BPB,Guo:MDSS}, and in combining sparse encoding (LDPC) methods with CDMA~\cite{Tanaka:SMALDPC}. Many of these papers however consider the {\it extreme dilution} regime -- in which the number of chip contributions is large but not $O(N)$.

The theoretical work regarding sparsely spread CDMA remained
lacking in certain respects. As pointed out in~\cite{Yoshida:ASS},
spreading codes with Poisson distributed number of non-zero
elements, per chip and across users, are systematically failing in
that each user has some probability of not contributing to any
chips (transmitting no information). Even in the ``partly
regular'' code~\cite{Montanari:ABP} ensemble (where each user
transmits on the same number of chips) some chips have no
contributors owing to the Poisson distribution in chip
connectivity, consequently the bandwidth is not effectively
utilised. We circumvent this problem by introducing constraints to
prevent this, namely taking regular signature codes constrained
such that both the number of users per chip and chips per user
take fixed integer values. Furthermore we present analytic and
numerical analysis without resort to Gaussian approximations of
any quantities. Using new tools from statistical mechanics we are
able to cast greater light on the nature of the binary prior
transmission process. Notably the nature of the decoding state
space and relative performance of sparse ensembles versus dense
ones across a range of noise levels; and importantly, the question
of how the coexistence of solutions found by
Tanaka~\cite{Tanaka:SMA} extends to sparse ensembles, especially
close to the transition points determined for the dense ensemble.

In this paper we demonstrate the superiority of regular sparsely
spread CDMA code over densely spread codes in certain respects,
for example, the anticipated bit error rate arising in decoding is
improved in the high noise regime and the solution coexistence
behaviour is less pervasive. Furthermore, to utilise belief
propagation for such an ensemble is certain to be significantly
faster and less computationally demanding~\cite{MacKay:ITI}, this also has 
power-consumption implications which may be important in some applications. Other practical issues
of implementation, the most basic being non-synchronisation and
power control, require detailed study and may make fully
harnessing these advantages more complex and application
dependent.

The paper is organised as follows: In section~\ref{model} we will
introduce the general framework and notation used, while the
methodology used for the various codes will be presented in
section~\ref{methodology}. The main results for the various codes
will be presented in section~\ref{results} followed by concluding
remarks in section~\ref{conclusion}.

%-------------------------------------------------------
\section{The model} \label{model}
%-------------------------------------------------------
\begin{figure}
  \begin{center}
    \setlength{\unitlength}{1cm}
    \begin{picture}(13.6, 5.2)(0.0,0.0)
      \put(0,0.2){\epsfbox{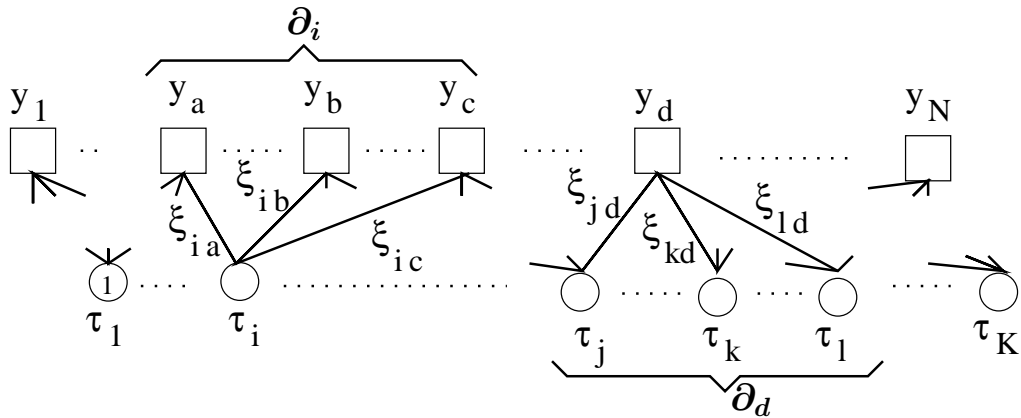}}
      \put(3.7,5.0){\large{\mbox{\boldmath{$\partial_i$}}}}
      \put(9.6,0.0){\large{\mbox{\boldmath{$\partial_d$}}}}
    \end{picture}
  \end{center}

  \caption{\label{bigraph} A bi-partite graph is useful for visually
  realising a problem. A user node $i$ at the bottom interacts with other variables
  through
  its set of neighbouring factor nodes ($\partial i$) to which it
  connects. The factor nodes are determined through a similar
  neighborhood. The interaction at each factor ($\mu$) is conditioned on neighbouring gain factors
 $\vxi^\mu$ (the non-zero components of ${\bf s}$), and $y_\mu$ (which is an implicit function of the
  noise $\omega_\mu$, and neighbouring input bits $\vb^\mu$ and gain factors $\vxi^\mu$), assuming a uniform prior on the bits. The statistical mechanics reconstruction
  problem associates dynamical variables $\vtau$ to the user nodes
  that interact through the factors.
The thermodynamical equilibrium state of this system then describes the theoretical
performance of optimal detectors.
}
\end{figure}

We consider a standard model of CDMA consisting of $K$ users
transmitting in a bit interval of $N$ chips. We assume a model
with perfect power control and synchronisation, and consider only
the single bit interval. In our case the received signal $\vy$ is
described by
\begin{equation}
  \vy =  \sum_{k=1}^K \left[\vs_k b_k\right] + \vomega\;,
\end{equation}
where the vector components describe the values for distinct
chips: $\vs_k$ is the spreading code for user $k$, $b_k=\pm 1$ is
the bit sent by user $k$ (binary input symbols) and $\vomega$ the
noise vector. Appropriate normalisation of the power is through
the definition of the signature matrix (${\bf s}$). It is possible
to include a user or chip specific amplitude variation, which may
be due to fading or imperfect power control. We consider a model
without these effects.
 The spreading codes are sparse so that
in expectation only $C$ of the elements in vector $\vs_k$ are non-zero.
If, with knowledge of the signature matrix in use, we assume the
signal has been subject to additive white Gaussian channel noise
of variance $\sigma_0^2/\beta$, where $\sigma_0^2$ is the variance
of the true channel noise $\langle\omega^2\rangle$, we can write
the posterior for the transmitted bits $\vtau$ (unknowns given the
particular instance) using Bayes Theorem
\begin{equation} 
\fl  P(\vtau | \vy) = \prod_{\mu=1}^N \left[ \frac{\sqrt{\beta}}{\sqrt{2
  \pi \sigma^2}} \exp\left(-\frac{\beta}{2 \sigma_0^2} \left(\sum_{k=1}^K
  \left[s_{\mu k} (b_k-\tau_k)\right] + \omega_\mu\right)^2
  \right)\right]P(\vtau)\;, \label{modelprob}
\end{equation}
and from this define bit error rate, mutual information, and other
quantities. The statistical mechanics approach from here is to
define a Hamiltonian and partition function from which the various
statistics relating to this probability distribution may be
determined - and hence all the usual information theory measures.
A suitable choice for the Hamiltonian is
\begin{equation}
  {\cal H}({\vtau}) = \sum_{\mu=1}^N \frac{1}{2 \sigma_0^2}\left(
  \sum_{k=1}^K \left[s_{\mu k} (b_k - \tau_k)\right] + \omega_\mu\right)^2 +
  \sum_{k=1}^K h_k \tau_k \;. \label{Ham}
\end{equation}
We can here identify ${\vtau}$ as the dynamical variables in the
inference problem (dependence shown explicitly). The other
quenched variables (parameters), describing the instance of the
disorder, are the signature matrix (${\bf s}$), noise ($\vomega$)
and the inputs ($\vb$).
The variables $h_k$ describe our prior beliefs about the inputs
(the specific user bias), and we can assume some simple distribution for this such as
all users having the same bias $h_k=H$. Maximal rate transmission corresponds
to unbiased bits $H=0$, and this is considered throughout the paper.
The properties of such a system may be
reflected in a factor (Tanner) graph, a bipartite graph in which
users and chips are represented by nodes (see figure~\ref{bigraph}).

The calculation we undertake is specific to the case of the
thermodynamic limit in which the number of chips $N \rightarrow
\infty$ whilst the load $\alpha=K/N$ is fixed. Note that $\alpha$
is termed $\beta$ in many CDMA papers, here we reserve
$\beta$ to mean the ``inverse temperature'' in a statistical
mechanics sense
(which defines our prior belief for the noise level and give rise
to the corresponding MAP detector.) 

In all ensembles we may identify the parameter $L$ as the mean
number of contributions to each chip, and $C$ as the mean number
of contributions per user. As such the following also holds
\begin{equation}
  \alpha= \frac{K}{N} = \frac{L}{C}\;.
\end{equation}
The case in which $\alpha$ is greater than 1 will be called
oversaturated, since more than one bit is being transmitted per
chip.

The calculations presented henceforth are specific to the case of
memoryless noise, drawn from a single distribution of mean zero and mean square $\sigma_0^2$
 \begin{equation} 
\Omega(\omega) = P(\omega_\mu=\omega)\;. 
\end{equation}
 Defining
normalised spreading codes such that $\sum_k \vs_k.\vs_k = N$
, we can identify the ``power spectral density'' ($PSD$) 
over a chip interval 
as a measure of the system
noise 
$1/(2 \sigma_0^2)$
-- the factor two being connected with physical considerations
in implementing the model.

\subsection{Code Ensembles}
\label{code_ensembles} We consider several code ensembles we call
irregular, partly regular and regular, which differ in the
constraints placed on the factor and variable degree constraints
of the signature matrix ${\bf s}$. The probability distribution
\begin{eqnarray}
P({\bf s}) &=& {\cal N} \left(\prod_\mu \left\langle \frac{\tL!}{L^\tL}  \delta(\sum_k \delta(s_{\mu k}\neq 0) - \tL)\right\rangle_{P(\tilde L)} \right)   \nonumber\\
&\times& \left(\prod_k\left\langle \frac{\tC!}{C^\tC}\delta(\sum_\mu \delta(s_{\mu k}\neq 0) - \tC) \right\rangle_{P(\tilde C)} \right)\prod_\mu \prod_k P(s_{\mu k}) \label{Ps}\;, 
\end{eqnarray}
where ${\cal N}$ is a normalising constant, $P(\tL)$ is the factor
degree probability distribution of mean $L$, $P(\tC)$ is the
variable degree probability distribution of mean $C$, and
$P(s_{\mu k})$ is the marginal probability distribution which is
common to all ensembles
\begin{equation}
 P(s_{\mu k})= \left(1-\frac{C}{N}\right)\delta(s_{\mu k}) +
 \frac{C}{N}\delta(s_{\mu k}-\xi)\label{PsMarg}\;.
\end{equation}
The form of (\ref{Ps}) is then sufficient for the sparse distributions we consider in the large system limit, and makes explicit the chip and user connectivity properties of the ensembles. 
The gain factor $\xi$, is drawn randomly from a single
distribution with zero measure at $\xi=0$, and finite moments, in
any instance of a code
\begin{equation}
 \phi(\xi) = P(s_{\mu k}=\xi |s_{\mu k} \neq 0)\;.
\end{equation}
Unlike the dense case the details of this distribution will effect
results, but only in a small way for reasonable choices~\cite{Yoshida:ASS}. 
We here investigate the case of Binary Phase
Shift Keying (BPSK) which corresponds to a uniform distribution on
$\{-\frac{1}{\sqrt{L}},\frac{1}{\sqrt{L}}\}$, though the analytic
results presented are applicable to any distribution of mean
square $=1/L$. Note that disorder in the gain factors is not a necessity, the
case $\xi=1/\sqrt{L}$ also allows decoding in sparse ensembles.

The case where $P(\tL)$ and $P(\tC)$ are Poissonian distributed
identifies the irregular ensemble - where the connections between
chips and users are independently distributed. The second
distribution called partly regular has $P(\tC)=\delta_{C,\tC}$, in
which the chip connectivity is again Poisson distributed with mean
$L$, but each user contributes to exactly $C$ chips. This prevents
the systematic failure inherent in the irregular ensemble since
therein an extensive number of users fail to transmit on any
chips. If in addition to the aforementioned constraint all chips
receive exactly $L$ contributions, $P(\tL)=\delta_{L,\tL}$, the
ensemble is called regular. Regular chip connectivity amongst
other things prevents the systematic inefficiency due to leaving
some chips unaccessed by any of the users. The case of Poissonian
distributions is that in which there is no global control. In many
engineering applications constraining users individually
(non-Poissonian $P(\tC)$) is practical, whereas coordination
between users (non-Poissonian $P(\tL)$) is difficult. The
practicalities of implementing the different ensembles we consider
are application specific: the advantages inherent in distributing
channel resources more evenly amongst users may be lost to
practical implentation problems.

%%%%%%%%%%%%%%%%%%%%%%%%%%%%%%%%%%%%%%%%%%%%%%%%%%%%%
\section{Methodology}
\label{methodology}
%%%%%%%%%%%%%%%%%%%%%%%%%%%%%%%%%%%%%%%%%%%%%%%%%%%%%%%

\subsection{Spectral Efficiency Lower Bound}
\label{icub}

The inferiority of codes with Poissonian user connectivity
 has been pointed out previously
(e.g., in~\cite{Yoshida:ASS}), based on the understanding that
codes which leave a portion of the users disconnected cannot be
optimal. Analogously we argue that codes with irregular chip connectivity
must also be inferior in that they leave a fraction of the chips (bandwidth) unutilised, 
thus providing a motivation for considering fully regular codes. 

In this section we 
show a particular case in which the regular codes are expected to outperform any other 
ensemble by analysing the amount of information that can be extracted
on the sent bits by consideration of only one chip in isolation of the other chips. This
corresponds to a detector reconstructing bits based only on the value of a
single chip, and is independent of the user connectivity. 

The spectral efficiency is defined as the mutual information
between the received signal and reconstructed bits per chip. In considering only a single chip ($\mu$) we have
\begin{equation}
I(\vtau;y_\mu)= \left\langle\log_2 \frac{P(\vtau | y_\mu)}{P(\vtau)}\right\rangle_{P_0(\vtau,y_\mu)}\;,
\end{equation}
where the subscript zero indicates that the true (generative),
rather than model (\ref{modelprob}), probability distribution. For brevity we consider the simplest case that the generative and model probability distributions are the same with unbiased bits and a Gaussian noise distribution in which case after some rearrangement
\begin{equation}
 I(\vtau;y_\mu)= \tL - \left\langle \log_2 \frac{\exp(-H_\mu(\vtau^\mu))}{\sum_{\vtau^\mu} \exp(-H_\mu(\vtau^\mu))} \right\rangle_{P_0(\vtau^\mu,y_\mu)} \label{MutInf_type1}\;,
\end{equation}
where $\vtau^\mu$ are the bits connected to chip $\mu$, and the chip Hamiltonian is
\begin{equation}
 H_\mu(\vtau^\mu) = \frac{1}{2\sigma_0^2} \left(-\sum_{i=1}^\tL \xi_i\tau_i + y_\mu \right)^2 \;, \label{Hamiltonian}
\end{equation}
 labelling each interacting (non-zero) component on the chip by $i$, $\tL$ being the chip connectivity. 

Working from this description we wish to compare the performance of ensembles with different chip connectivities. To do this we consider the ensemble average mutual information by averaging the mutual information over the connectivities ($\tL$), load factors, and transmitted bits. This average is complicated, however it is possible to calculate the dominant terms in the low and high $PSD$ limits. 

In the case of low noise ($PSD\rightarrow\infty$) we find the asymptotically dominant terms come first from the numerator
\begin{equation}
 \left\langle\log_2 \exp - H(\vtau^\mu)\right\rangle \doteq \left\langle\frac{\omega^2}{2\sigma_0^2}\right\rangle/\log(2) = \frac{1}{2\log(2)}\label{energeticpart}\;,
\end{equation}
which is an average over the ground state energy, and also the logarithm of the denominator which is
\begin{equation}
\fl \left\langle\log_2 \sum_{\vtau^\mu} \exp - H(\vtau^\mu) \right\rangle \doteq \left\langle\log_2 \left[ \sum_{\vtau^\mu}\exp\left(\frac{-\omega^2}{2\sigma_0^2}\right)\delta\left(\sum_i \xi_i(b_i -\tau_i)\right)\right]\right\rangle \label{entropicpart}\;,
\end{equation}
where $y_\mu$ has been decomposed into its bit ($\{b_i\}$) and noise ($\omega$) parts, and the averages are now over the ensembles as well as $y_\mu$. The first part of (\ref{entropicpart}) gives an energy contribution cancelling (\ref{energeticpart}). We call the remaining part the average over the {\it chip entropy}, by comparison with (\ref{MutInf_type1}) this determines the amount of information lost in decoding. The chip entropy term contains an indicator function counting the ground states - the average chip entropy is zero when $\vtau^\mu=\vb^\mu$ is the only solution. For the case of $BPSK$ however there may be some degeneracy in ground states with two terms in the sum being non-zero but cancelling one another. This degeneracy has a dependence on the distribution $P(\tL)$ for given $L$. Averaging over load factors and transmitted bits we find that in the zero noise limit
\begin{eqnarray}
 I(\vtau, y_\mu) &\doteq& L - \left\langle \frac{1}{2^{2\tL}} \sum_{\vxi^\mu} \sum_{\vb^\mu} \log_2 \sum_{\vtau^\mu}\delta\left(\sum_i \xi_i(b_i -\tau_i)\right)\right\rangle_{P(\tL)}\;,\\
&=& L - \left\langle \sum_{p=0}^\tL \frac{1}{2^\tL} {\tL \choose p} \ln \left(\sum_i^{\min(p,\tL-p)} {\tL - p \choose i} {p \choose i} \right)\right\rangle_{P(\tL)} \label{entropylowerbound}\;.
\end{eqnarray}
By numerical evaluation of this function (see results section \ref{selbnr}) we find that the optimal ensemble is in fact the regular ensemble. This is because chip entropy, when averaged over load factors and bits is a concave function in $\tL$, so that the information loss is minimised when $P(\tL)=\delta_{L,\tL}$. This dependency on $\tL$ may be a peculiarity of the detector considered, but many other aspects of the calculation may be generalised to give a similar result.

It is possible to consider the opposite limit $\sigma_0^2
\rightarrow \infty$ perturbatively. We found that the leading four
orders in $1/{\sigma_0}$ were identical for all code ensembles of
the same mean chip connectivity. We would anticipate the behaviour at 
non-extreme $PSD$ to fall somewhere between these two regimes and 
thus for the chip regular ensemble to be atleast as good as the chip irregular ensembles.

We note here that another reason for considering the regular code optimal amongst sparse random codes is to consider the field term when the Hamiltonian (\ref{Hamiltonian}) is written in canonical form with a set of couplings ($\{J_{\langle i j\rangle}\}$) and user specific external fields ($\{h_i\}$). In this representation the set of external fields are in expectation aligned with the sent bit sequence, but subject to fluctuations for each code instance. The variance of these fluctuations may be shown to be proportional to the excess chip connectivity over the true chip connectivity~\cite{Raymond:RM}, which amongst all ensembles is minimised by the regular chip ensemble. The multi-user interference is larger in irregular codes and hence information recovery is weaker as predicted in this section.\footnote[1]{This argument is added since published version.}

\subsection{Replica Method Outline}
\label{rmo}
We determine the static properties of our model defined in section \ref{model}, including
correlations due to the full interaction structure, we use the replica method. 
 From
the expression of the Hamiltonian (\ref{Ham}) we may identify a
free energy and partition function as:
\begin{equation*}
  f = -\frac{1}{N\beta} \ln  Z \qquad\qquad
  Z = \Tr_{{\vtau}} \exp \left(-\beta {\cal H}({\vtau})\right)\;.
\end{equation*}

To progress we make use of the anticipated {\it self-averaging}
properties of the system. The assumption being that in the large
system limit any two randomly selected instances will, with high
probability, have indistinguishable statistical properties. This
assumption has firm foundation in several related
problems~\cite{Vicente:LDPC}, and is furthermore intuitive after
some reflection. If this assumption is true then the statists of
any particular instance can be described completely by the free
energy averaged over all instances of the disorder. We are thus
interested in the quantity
\begin{equation}
  {\cal F} = \langle f \rangle = -\lim_{N\rightarrow\infty}\frac{1}{N\beta}
  \langle \ln  Z \rangle_I\;, \label{freeenergy1}
\end{equation}
where the angled brackets represent the weighted averages over $I$
(the instances). 
The entropy density may be calculated from the free energy density
by use of the relation
\begin{equation}
  s=\beta(e-f) \ ,
\end{equation}

where $e$ is the energy density.

To determine the free energy we must average over disorder in (\ref{freeenergy1}), which is a difficult problem except in special cases.
This is why we make use of the replica identity
\begin{equation}
  \langle \ln  Z \rangle_I = \lim_{n \rightarrow 0} \frac{\partial}{\partial n} \langle Z^n
  \rangle_I\;. \label{rep_part_fun}
\end{equation}
We can model the system now as one of interacting replicas, where
$Z^n$ is decomposed as a product of an integer number of partition
functions with conditionally independent (given the instance of
the disorder) dynamical variables. The discreteness of replicas is
essential in the first part of the calculation, but a continuation
to the real numbers is required in taking $n\rightarrow0^{+}$ --
this is a notorious assumption, which rigorous mathematics can not
yet justify for the general case, in spite of the progress made in
recent years~\cite{Talagrand:GP,Franz:RB,Guerra:BRS}. However, we shall assume
validity and since the methodology for the sparse structures is
well established~\cite{Monasson:OP,Wong:GB,Vicente:LDPC} we omit our 
particular details. The final functional form for the free energy
is determlained from
\numparts
\begin{eqnarray}
\fl \langle Z^n \rangle &=& \int \prod_{\vsigma,b}
\left[dP(b,\vsigma)d{\hat P}(b,\vsigma)\right] \exp\{\ln{\cal N} +
N(G_1(n) + G_2(n) +G_3(n)) \} \;;\\
\fl  G_1(n) &=& \ln \left\{ \int \right[\prod_\alpha
  \frac{\rmd\lambda_\alpha}{\sqrt{2\pi}}\left] \exp \left\{-\sum_\alpha
\lambda_\alpha^2/2\right\}\left\langle \exp\left\{
\frac{\rmi\sqrt{\beta}\omega}{\sigma_0}\sum_\alpha
\lambda_\alpha\right\}\right\rangle_{\Omega(\omega)} \right. \nonumber\\
\fl    &\times& \left. \left\langle e^{-L}
    \left(\sum_{b,\vsigma} P(b,\vsigma) \left\langle
    \exp\left\{\frac{\rmi\sqrt{\beta}\xi}{\sigma_0}
    \sum_\alpha \lambda_\alpha (b - \tau_\alpha)\right\}
    \right\rangle_{\phi(\xi)} \right)^\tL \right\rangle_{P(\tilde L)} \right\} \;;\\
 \fl G_2(n) &=& \sum_{\vsigma,b} P(b,\vsigma){\hat P}(b,\vsigma) \;;\\
 \fl G_3(n) &=& \alpha \ln \left\langle \sum_{\vtau} \exp\left\{\beta H
\sum_\alpha \tau_\alpha\right\} \left\langle \frac{1}{(-L)^{\tC}}\left({\hat P}(b,\vtau)
\right)^{\tC}\right\rangle_{P(\tC)}\right\rangle_{P_0(b)} \;;\label{G3full}
\end{eqnarray}
\endnumparts
where ${\cal N}$ is a constant due to normalising the ensembles
(\ref{Ps}). This expression may be evaluated at the saddle point
to give an expression for the free energy. In the term
(\ref{G3full}) we account for the cases in which the marginalised
probability distribution $P_0(b)$ and assumed marginal probability
distribution (described by $H$) are asymmetric. In the case of
maximal rate which we will consider, the $b$ average is trivial
and $H=0$. Provided that in addition the
gain factor distribution is symmetric then it is
possible to remove the $b$ dependence in the order parameters,
since the symmetry $P(b,\vsigma)=P(-b,-\vsigma)$ and  ${\hat
P}(b,\vsigma)={\hat P}(-b,-\vsigma)$ leaves the free energy
invariant.

\subsection{Replica Symmetric Equations}
The concise form for our equations is attained using the
assumption of replica symmetry (RS). This amounts to the
assumption that the correlations amongst replicas are all
identical, and determined by a unique shared distribution. The
validity of this assumption may be self consistently tested
(section \ref{Stability}). This assumption differs from that used
by Yoshida and Tanaka~\cite{Yoshida:ASS} where the correlations
are described by only a handful of parameters rather than a
distribution once RS is assumed -- this approach may therefore
miss some of the detailed structure although it is easier to
handle numerically. The order parameter in our case is given by
\numparts
\begin{eqnarray}
  P(b,\vtau) = \frac{1}{2}\int d\pi(x) \prod_\alpha \left(\frac{1}{2}(1+b\tau_\alpha x) \right)\;;\\
 {\hat P}(b,\vtau) = {\hat q} \int d{\hat \pi}({\hat x}) \prod_\alpha \left(1 + b\tau_\alpha x \right) \;;
\end{eqnarray}
\endnumparts
where ${\hat q}$ is a variational normalisation constant and
$\pi,{\hat \pi}$ are normalised distributions on the interval
$[-1,1]$. From here onwards we may consider the case in which the
bit variables $\tau_\alpha$ and gain factors $\xi$ are
gauged to $b$ ($\tau b \rightarrow \tau$, $\xi b \rightarrow
\xi$).

Using Laplace's method, this gives the
following expression for the (RS) free energy at the saddle point
\begin{equation}
  {\cal F}_{RS} = -\frac{1}{\beta} \mbox{Extr}_{\pi,\widehat{\pi}} \frac{\partial}{\partial n}\left( {\cal G}_{1,RS}(\tL)(n) + {\cal G}_{2,RS}(n) + {\cal G}_{3,RS}(\tC)(n) \right) \label{freeenergy}
\end{equation}
where
\numparts
\begin{eqnarray}
\fl  \left. \frac{\partial}{\partial n} \right|_{n=0} {\cal
G}_{1,RS}(n) &\!\doteq\!&  -L\ln 2 \nonumber \\
&\!+\!& \left\langle\int
\!\prod_{l=1}^\tL
    \left[\rmd\pi(x_l)\right] \left\langle\ln \mbox{Tr}_{\{\tau_l=\pm 1\}}
    \chi_\tL(\vtau;\{\xi\},\omega,\{x\}) \right\rangle_{\Omega(\omega),\phi(\xi)} \right\rangle_{P(\tL)}
    \;;\label{G1}\\
\fl \chi_\tL(\vtau;\{\xi\},\omega,\{x\}) &=&
    \exp\left(-\frac{\beta}{2
      \sigma^2}\left(\omega+\sum_{l=1}^\tL
    (1-\tau_l)\xi_l\right)^2\right) \prod_{l=1}^\tL (1+\tau_l x_l)\label{singlenodeinter}
    \;; \\
 \fl    \left.  \frac{\partial}{\partial n}\right|_{n=0} {\cal G}_{2,RS}(n) &\!=\!& - L \int \rmd\pi(x_c) \rmd{\hat \pi}({\hat x}_c) \ln (1 + x
      {\hat x}_c)
      \;;\\
\fl \left. \frac{\partial}{\partial n}\right|_{n=0} {\cal
G}_{3,RS}(n) &\!=\!& \alpha \left\langle\int \prod_{c=1}^\tC
\left[\rmd{\hat
      \pi}({\hat x}_c)\right] \ln \left(\prod_{c=1}^\tC (1+{\hat x}_c) + \prod_{c=1}^\tC
      (1-{\hat x}_c) \right)\right\rangle_{P(\tC)}\;.\label{G3}
\end{eqnarray}
\endnumparts
and the saddle point value for ${\hat w}$ ($=L$) has been introduced.
The averages over $\tL$ and $\tC$ encapsulate the differences amongst the ensembles.

Equation (\ref{singlenodeinter}) describes the interaction at a
single chip in the factor graph (figure \ref{bigraph}) of
connectivity $\tL$. The parameter $\xi_l$ and variable $\tau$ are
the gain factors, and reconstructed bits respectively,
both gauged to the transmitted bit, while $\omega$ is the instance
of the chip noise.

The order variational distributions $\{\pi,\hat{\pi}\}$ are chosen
so as to extremise~(\ref{freeenergy}).  The self consistent equations attained
by the saddle point method are:
\numparts
\begin{eqnarray}
\fl{\hat \pi}({\hat x}) &=& \left\langle\int \prod_{l=1}^{\tL} \left[\rmd \pi(x_l)\right] \left<\delta\left({\hat x} \!-\! \frac{\mbox{Tr}_{\{\tau_l=\pm 1\}} ~\tau_{\tL+1} ~{\bar \chi}_{\tL}(\vtau;\{\xi\},\{{\hat x}\}) }
{\mbox{Tr}_{\{\tau_l=\pm 1\}} ~{\bar \chi}_{\tL}(\vtau;\{\xi\},\omega,\{x\})}\right)
\right>_{\{\xi\},\omega}\right\rangle_{P(\tL)}\; \label{saddlepoint1}\\
\fl{\bar \chi}_\tL(\vtau;\{\xi\},\omega,\{x\}) &=&
\exp\left(-\frac{\beta}{2 \sigma^2}\left(\omega+\sum_{l=1}^{\tL+1} (1-\tau_l)\xi_l\right)^2\right) \prod_{l=1}^{\tL}(1+\tau_l x_l) \\
 \fl\pi (x) &=& \left\langle \int \prod_{c=1}^{\tC} \left[\rmd{\hat \pi}({\hat x}_c)\right] \delta\left(x - \frac{\prod_{c=1}^{\tC} (1+{\hat x}_c) - \prod_{c=1}^{\tC} (1-{\hat x}_c)}{\prod_{c=1}^{\tC} (1+{\hat x}_c) + \prod_{c=1}^{\tC} (1-{\hat x}_c)} \right)\right\rangle_{P(\tC)} \;.\label{saddlepoint2} \;
\end{eqnarray}
\endnumparts
The variables $P(\tL)$ and $P(\tC)$ are here the excess degree
distributions of the particular ensemble (\ref{Ps}). For regularly
constrained ensembles the chip and user excesses are $L-1$ and
$C-1$ respectively. For Poissonian distributions the excess degree
distribution is the full degree distribution.

Aside from entropy, the other quantities of interest may be determined from the
probability distribution for the overlap of reconstructed and sent
variables $m_k=\langle \tau_k \rangle$,
\begin{eqnarray}
  P (m) &=& \lim_{K\rightarrow\infty}\frac{1}{K} \left\langle\sum_{k=1}^K \delta_{m
_k,m}\right\rangle_I \;,\\
	&=&
  \left\langle \int \prod_{c=1}^{\tC} \left[\rmd {\hat \pi}({\hat x}_c)\right]
  \delta\left(m \!-\! \frac{\prod_{c=1}^{\tC} (1+{\hat x}_c) \!-\! \prod_{c=1}^{\tC}
  (1-{\hat x}_c)}{\prod_{c=1}^{\tC} (1+{\hat x}_c) \!+\! \prod_{c=1}^{\tC}
  (1-{\hat x}_c)} \right)\right\rangle_{P(\tC)}\;.\\
\end{eqnarray}

We note finally that equivalent expressions to these found with the RS assumption
may be obtained by using the cavity method~\cite{Mezard:SGT} with the assumption of a single pure state. 
This approach is a probabilistic one and hence more intuitive on some levels. 

\subsection{Population Dynamics}
Analysis of these equations is primarily constrained by the nature of
equations (\ref{saddlepoint1}-\ref{saddlepoint2}). No exact solutions are apparent,
and perturbative regimes about the ferromagnetic solution (which
is only a solution for zero noise) are difficult to handle.
Consequently we use population dynamics~\cite{Mezard:BLSG} --
representing the distributions $\{\pi(x) , {\hat
\pi}({\hat x})\}$ by finite populations (histograms) and iterating
this distribution until convergence. It
is hoped, and observed, that each histogram captures sufficient detail to
 describe the continuous function and the dynamics (described below) allow 
convergence towards a true solution distribution with only 
small corrections due to finite size effects.

To solve the equations (\ref{saddlepoint1},\ref{saddlepoint2}) with population dynamics finite histograms constucted from $M$ undirected {\it cavity} magnetisations are used. Histograms approximating each function are formed
\numparts
\begin{eqnarray}
 \pi(x) \rightarrow W = \{x_1,\ldots,x_i,\ldots,x_M\}\;, \label{phist}\\
 \pi({\hat x}) \rightarrow {\hat W} = \{{\hat x}_1,\ldots,{\hat x}_a,\ldots,{\hat x}_M\}\label{phathist}\;,
\end{eqnarray}
\endnumparts
with $M$ sufficiently large to provide good resolution in the desired performance measures. 
The discrete minimisation dynamics of the histograms is derived from (\ref{saddlepoint1}-\ref{saddlepoint2}). Histogram updates are undertaken alternately, with all magnetisation in the histogram being updated sequentially. In the update of field $x_a$ the quenched parameters $\{\tL, \omega, \vxi\}$ are sampled, $\tL$ being the chip excess degree, and $\tL$ magnetisations are randomly chosen from $W$, defining through (\ref{saddlepoint1}) the update
\begin{equation}
  {\hat x}_a = \frac{\mbox{Tr}_{\{\tau_l=\pm 1\}} ~\tau_{\tL+1} ~{\bar \chi}_{\tL}(\vtau;\{\xi\},\omega,\{x\}) }
{\mbox{Tr}_{\{\tau_l=\pm 1\}} ~{\bar \chi}_{\tL}(\vtau;\{\xi\},\omega,,\{x\})} \;. \label{hist1}
\end{equation}
 The update of the other histogram follows dynamics in which $\tC$ is sampled, $\tC$ being the user excess degree, along with $\tC$ randomly chosen magnetisations from ${\hat W}$, defining through (\ref{saddlepoint2}) the update
\begin{equation}
 x_i = \frac{\prod_{c=1}^{\tC} (1+{\hat
x}_c) - \prod_{c=1}^{\tC} (1-{\hat x}_c)}{\prod_{c=1}^{\tC}
(1+{\hat x}_c) + \prod_{c=1}^{\tC} (1-{\hat x}_c)}\;.\label{hist2}
\end{equation}

There is a strong analogy between the population dynamics
algorithm and that of message passing on a particular instance of
the graph. The iteration of the histograms implicit in
(\ref{hist1}-\ref{hist2}) is analogous to the
propagation of a population of {\it cavity} magnetisations between
factor ($a$) and user ($i$) nodes, which may be written as the
self consistent equations: 
\numparts
\begin{eqnarray} 
    {\hat x}_{a\rightarrow i} &=&
   \frac{1}{{\cal N}_{\hat{x}}} \mbox{Tr}_{\{\tau_l=\pm 1\}} \tau_i \exp\left(-\frac{\beta}{2 \sigma^2}\left(\omega_a + \sum_{l  \in \partial a \setminus i} (1-\tau_l)\xi_{al}\right)^2\right)  \nonumber \\ &\times& \prod_{l\in
  \partial a \setminus i} (1+\tau_l x_{l\rightarrow a}) \label{cavity}
  \;; \\
  x_{i \rightarrow a} &=& \frac{1}{{\cal N}_x} \left(\prod_{c\in \partial i
  \setminus a} (1+{\hat x}_{c\rightarrow i}) - \prod_{c\in \partial i \setminus a} (1-{\hat x}_{c\rightarrow i})\right)\;; \label{cavity2}
\end{eqnarray}
\endnumparts
where ${\cal N}_{x,\hat{x}}$ are the relevant normalisations, and
the abbreviation ${\partial y}$ indicates the set of nodes
connected to $y$. In population dynamics, the notion of a
particular graph with labelled edges is absent however, and the
only the distribution of the two types of magnetisations are
relevant. 

\subsection{Stability Analysis}
\label{Stability}
To test the stability of the obtained solutions we consider both the
appearance of non-negative entropy, and a stability parameter
 defined through a consideration of the fluctuation
dissipation theorem. The first criteria that the entropy be
non-negative is based on the fact that physically viable solutions in
discrete systems must have non-negative entropy so that any solution found not
meeting this criteria must be based on bad premises; replica symmetry
 is a likely source.

The stability parameter $\lambda$ is defined in connection with
the cavity method for spin glasses~\cite{Rivoire:GM} and tests
local stability of the solutions.
It is equivalent to testing the local stability of belief
propagation equations as proposed in~\cite{Kabashima:PB}. A
necessary condition for the stability of the $RS$ solution is that
the corresponding susceptibility does not
diverge. This condition ensures that fields are not strongly
correlated. The spin glass susceptibility when averaged over
instances may be defined
\begin{equation}
 \zeta = \sum_{d=0}^\infty  X^d \left\langle \left\langle \tau_0 \tau_d \right\rangle_c^2 \right\rangle\;, \label{suscept}
\end{equation}
where $d$ is the distance between two nodes in the factor graph,
the inner average denotes the connected correlation function
between these nodes, $X^d$ describes the typical number of
variables at distance $d$, and the outer average is over instances
of the disorder (self-averaging part). This quantity is not
divergent provided that
\begin{equation}
 \lambda = \ln \left[\lim_{d\rightarrow\infty} X \left\langle \left\langle \tau_0 \tau_d \right\rangle_c^2 \right\rangle^{\frac{1}{d}}\right]
\end{equation}
is negative, since this indicates an asympoptically exponential
decrease in the terms of (\ref{suscept}) and hence convergence of
the sum. In the thermodynamic limit the connected correlation
function is dominated by a single direct path which may be
decomposed as a chain of local linear susceptibilities
\begin{eqnarray}
 \left\langle\tau_0 \tau_d\right\rangle_c \propto  \prod_{(i,j)} \frac{\partial x_{i\rightarrow a}}{\partial {\hat
  x}_{b\rightarrow i}} \frac{\partial {\hat x}_{b\rightarrow i}}{\partial x_{j\rightarrow b}}\;,
\end{eqnarray}
where (i,j) indicate the set of variables on the shortest path
between nodes $0$ and $d$ in a particular instance of the graph
(\ref{cavity}). 

This representation allows us to construct an
estimation for $\lambda$ numerically based on principles similar
to population dynamics~\cite{Raymond:PD} -- the directedness and fixed structure implicit in a particular problem is removed with the self-averaging assumption leaving a functional description similar to (\ref{saddlepoint1}-\ref{saddlepoint2}), which may be iterated.
In order to approximate the stability parameter $\lambda$ one introduces additional positive 
numbers in the population dynamics histograms (\ref{phathist},\ref{phist}),
$x_i\rightarrow\{x_i,v_i\}$ and ${\hat x}_a\rightarrow\{{\hat x}_a,{\hat v}_a\}$ respectively.
These new values represent the relative sizes of perturbations in each magnetisation, 
 and are updated in parallel to (\ref{hist1},\ref{hist2}) as 
\begin{equation}
 {\hat v}_a = \sum_j^\tL v_j \left(\frac{\partial {\hat x}_a}{\partial x_j}\right)^2\;,
\end{equation}
and with similar assignments for the field update of $W$
\begin{equation}
 v_i = \sum_j^\tC {\hat v}_a \left(\frac{\partial x_i}{\partial {\hat x}_a}\right)^2\;.
\end{equation}
The partial derivatives are calculated from (\ref{hist1}-\ref{hist2}) and evaluated at the corresponding values in the sampled population.
If the final fixed point is stable against small perturbations in the initial field then these values $\{v, {\hat v}\}$ must decay exponentially on average. Renormalisation of $\{v_i\}$ and $\{{\hat v}_a\}$ such that the mean is $1$ after each update is necessary. The numerical renormalisation constant for each population yields (dependent) estimations of $\lambda$, which can be sampled at a suitable convergence time (end of the $\{W,{\hat W}\}$ minimisation process). 

Like population dynamics we expect behaviour to be sensitive to initialisation 
conditions and finite size effects in some circumstances. 
In addition the estimation 
requires good resolution in the histograms $W$ and ${\hat W}$.

%-------------------------------------------------------
\section{Results}
\label{results}
%-------------------------------------------------------
Results are presented here for the canonical case of Binary Phase
Shift Keying (BPSK) where $\xi_l \in \{1,-1\}$ with equal
probability. Furthermore, we assume an AWGN model for the true
noise $\omega$ (of variance $\sigma_0^2$). For evaluation purposes
we assume the channel noise level is known precisely, so that
$\beta=1$, employing the Nishimori
temperature~\cite{Nishimori:SP}. This guarantees that the RS
solution is thermodynamically dominant. Furthermore the energy
takes a constant value at the Nishimori temperature and hence the
entropy is affine to the free energy. Where of interest we plot
the comparable statistics for the Single User Gaussian channel
(SUG), and the densely spread ensemble, each with MPM detectors --
equivalent to maximum likelihood for individual bits.

For population dynamics two parallel populations (\ref{phist},\ref{phathist}) are initialised either uniformly at
random, or in the ferromagnetic state. These two populations are
known to converge towards the unique solution, where one exists,
from opposite directions, and so we can use their convergence as a
criteria for halting the algorithm and testing for the appearance
of multiple solutions. In the case where they converge to
different solutions we can usually identify the solution converged to from
the ferromagnetic initial state as a {\it good} solution - in the sense
that it reconstructs well, and that arrived at from random
initial state as a {\it bad} solution. In the equivalent belief
propagation algorithm one cannot choose initial conditions equivalent to
ferromagnetic -- knowing the exact solution would of course makes
the decoding redundant. We therefore expect the properties of the
{\it bad} solution to be those realisable by belief propagation
(though clever algorithms may be able to escape to the good
solution under some circumstances). The stability variables $\{v,{\hat v}\}$ 
were initialised independently each as the square of a value drawn from a gaussian distribution -- and tests indicated other reasonable distributions produced similar results.

Computer resources restrict the cases studied in detail to
an intermediate $PSD$ regime, and small $L$.  In particular, the
problem at low $PSD$, is the Gaussian noise average, which is
poorly estimated, while at high $PSD$ a majority of the histogram
is concentrated at magnetisations $x,{\hat x}\approx 1$ not
allowing sufficient resolution in the rest of the histogram.

Several different measures are calculated from the converged order
parameter, indicating the performance of sparsely-spread CDMA.
Using the converged histograms for the fields we are able to
determine the following quantities: free energy, energy and a
histogram for the probability distribution, from discretisations
of the previously presented equations
(\ref{saddlepoint1}-\ref{saddlepoint2}). Using the probability
distribution we are also able to approximate the decoding bit
error rate
\begin{equation}
  P_b = \int \rmd P(m) \frac{1-\sign(m)}{2}\;; \\
\end{equation}
multi-user efficiency
\begin{equation}
  MuE = \frac{1}{SNR} \left[ \erfc^{-1}(P_b) \right]^2\;;
\end{equation}
and mutual information between sent and reconstructed bits per
chip, $I(\vb;\vtau)/N$ (taking a factorised form given the RS
assumption)
\begin{equation}
  MI = \alpha \left( 1 - \int \rmd P(m) \sum_{\tau} \frac{1+\tau m}{2} \log_2\frac{1+\tau m}{2} \right) \;.\\
\end{equation}
The spectral efficiency is the capacity $I(\vtau;\vy)$ per chip,
which is affine to the entropy (and the free energy at the Nishimori temperature)
\begin{equation}
  \nu = \alpha - s/\ln 2\;.\\
\end{equation}
Negative entropy can be identified when the measured spectral
efficiency exceeds the load, and thermodynamic transition points
correspond to points of coincident spectral efficiency.

\begin{figure}
  \begin{center}
    \leavevmode
    \epsfxsize=7.5cm
    \epsfbox{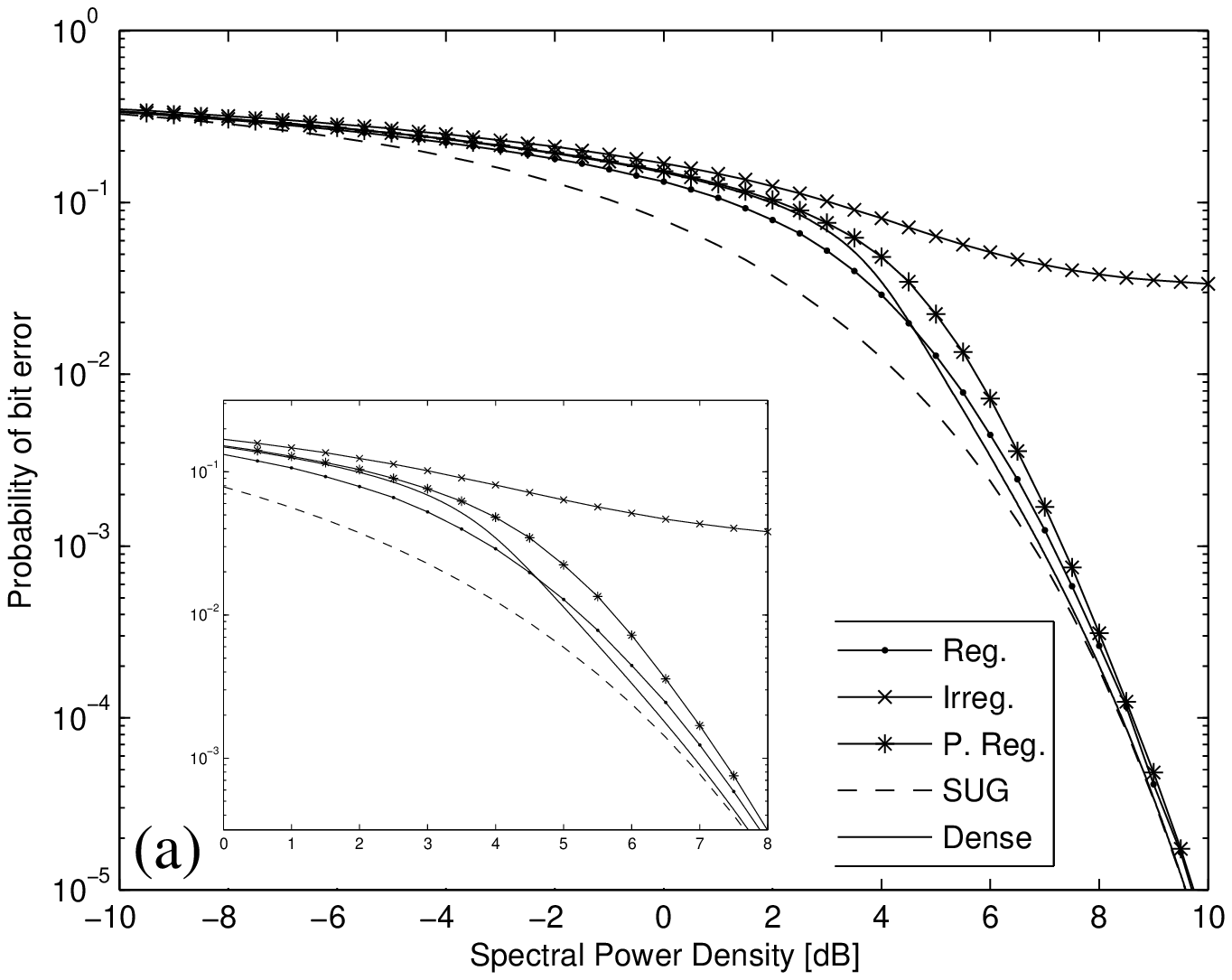}
    \epsfxsize=7.5cm
    \epsfbox{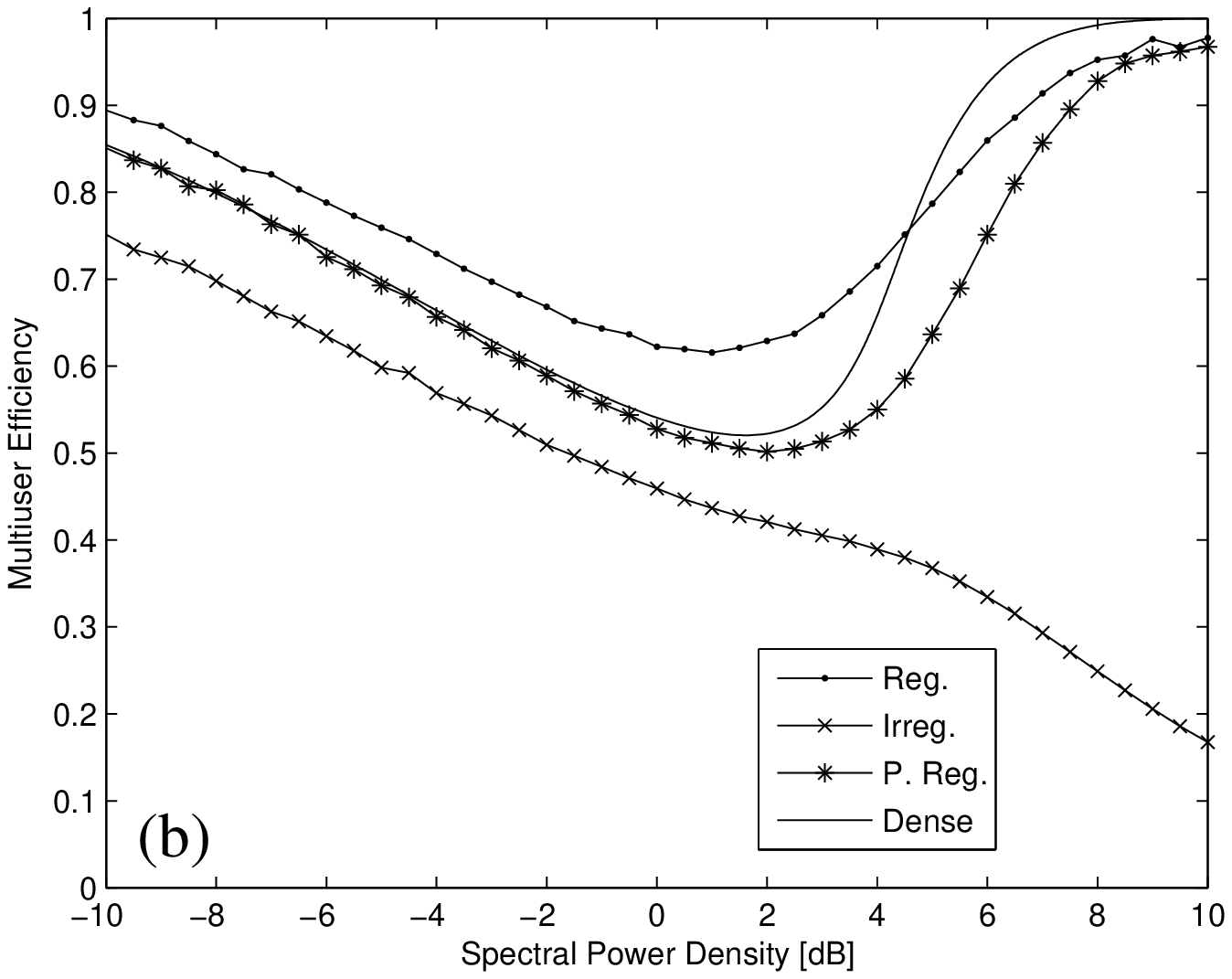}
    \epsfxsize=7.5cm
    \epsfbox{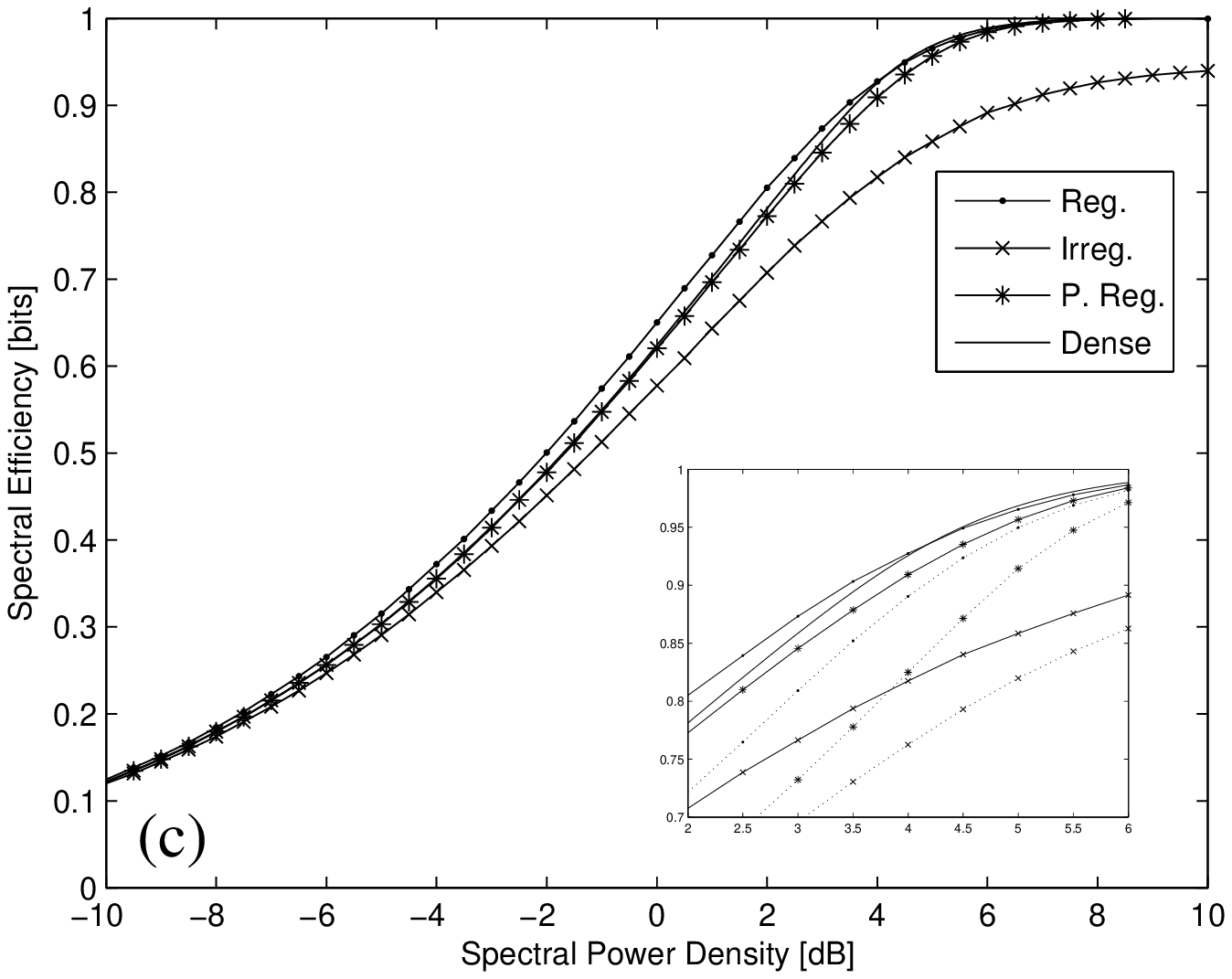}
    \epsfxsize=7.5cm
    \epsfbox{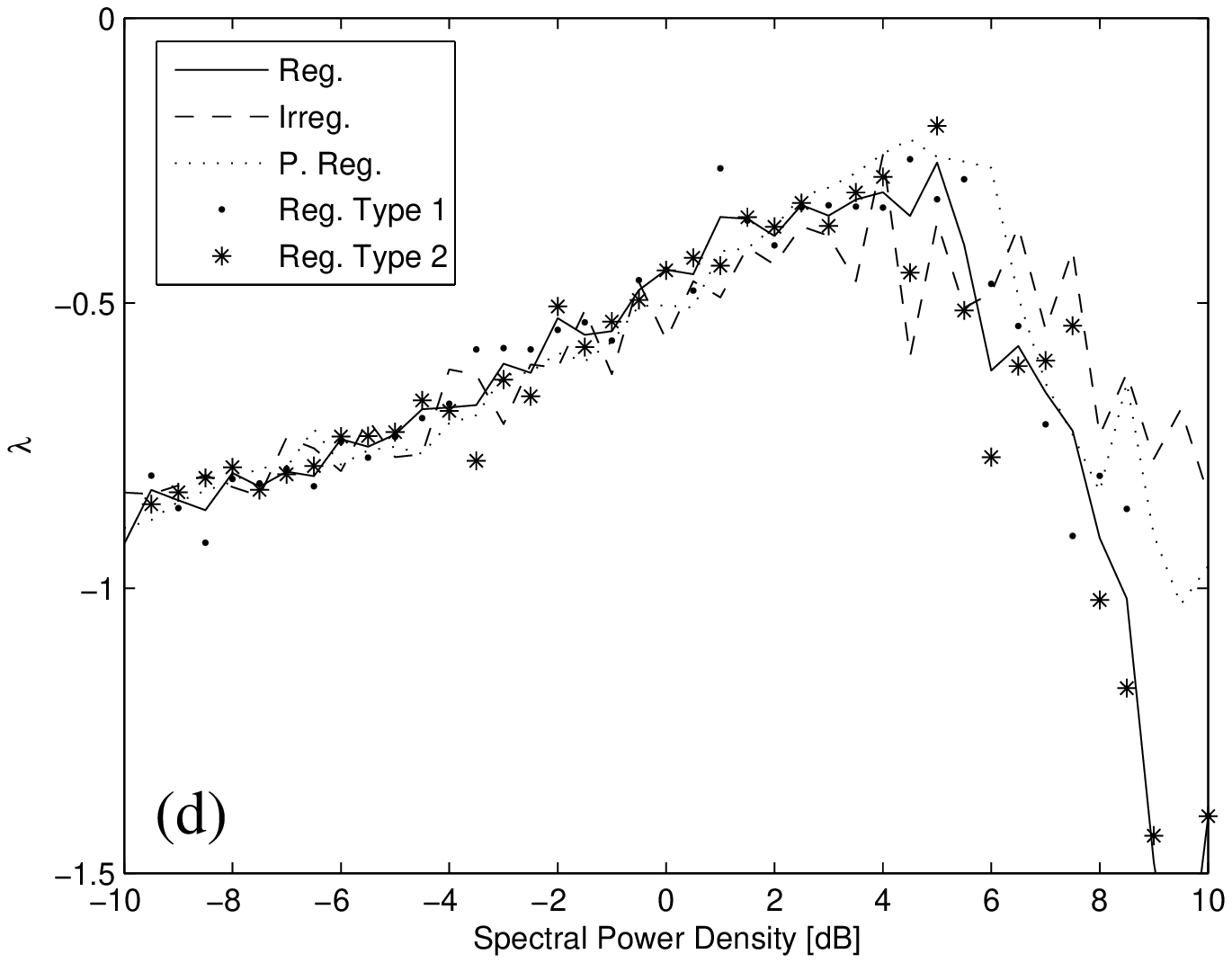}
\end{center}
\caption{\label{general_prop}  Performance of the sparse CDMA
configuration of variable and factor degree connectivities
$C:L=3:3$, respectively; all data presented on the basis of 100
runs, error bars are omitted and are typically small in subfigures
(a)-(c) the smoothness of the curves being characteristic of this
level (numerical accuracy was excellent only at intermediate
$PSD$s). (a) The bit error rate is limited by the disconnected
component in the case of irregular codes, otherwise trends match
the dense case, lower bounded by the SUG.  Inset - the range where
the sparse-regular and dense cases crossover.(b) Multiuser
efficiency indicates the regular user connectivity codes
outperform the dense case below some $PSD$. (c) The spectral
efficiency [$\full$] demonstrates similar trends, the entropy
being positive. The gap between the mutual information [$\dotted$]
and spectral efficiency (shown in the inset) is everywhere small
and especially so at small and large $PSD$, indicating little
information loss in the decoding process. (d) The two markers show
the mean results for the two different stability estimates in the
algorithm for the regular code. There are systematic errors at
small $PSD$, and convergence is good only at intermediate $PSD$.
The lines represent the average of these quantities for each
ensemble -- all ensembles show a cusp at some $PSD$, for $3:3$
codes the various ensembles shows very similar trends, indicating
local stability everywhere.}
\end{figure}

Figure~\ref{general_prop}\footnote[2]{This figure has been modified from the published version, the difference being that the Poissonian chip connectivity codes have everywhere weaker performance than the dense and sparse regular code ensemble.} demonstrates some general properties of
the regular ensemble in which the variable and factor degree
connectivities are $C:L = 3:3$, respectively.
Equations~(\ref{saddlepoint1}-\ref{saddlepoint2}) were iterated
using population dynamics and the relevant properties were
calculated using the obtained solutions; the data presented is
averaged over 100 runs and error-bars, which are typically small,
are omitted for brevity. Figure~\ref{general_prop}(a) shows the
bit error rate in regular and Poissonian codes, the inset focuses
on the range where the sparse-regular and dense cases crossover.
The sparse codes demonstrate similar trends to the dense case
except the irregular code, which show weaker performance in
general, and in particular at high $PSD$. Detailed
trends can be seen in figure~\ref{general_prop}(b) that shows the
multiuser efficiency. Codes with regular {\it user} connectivity show
superior performance with respect to the dense case at low $PSD$.
Figure~\ref{general_prop}(c) shows similar trends in the spectral
efficiency and mutual information (shown in the inset); the effect
of the disconnected (user) component is clear in the fact that the
irregular code fails to reach capacity at high noise levels. In
general it appears the chip connectivity distribution is not
critical in changing the trends present, unlike the user
connectivity distribution. It was found in these cases (and all
cases with unique solutions for given $PSD$), that the algorithm
converged to non-negative entropy values and to a stability
measure fluctuating about a value less than $0$, as shown
in figure~\ref{general_prop}(d). These points would indicate the
suitability of the RS assumption.

\begin{figure}
  \begin{center}
    \leavevmode
    \epsfxsize=7.5cm
    \epsfbox{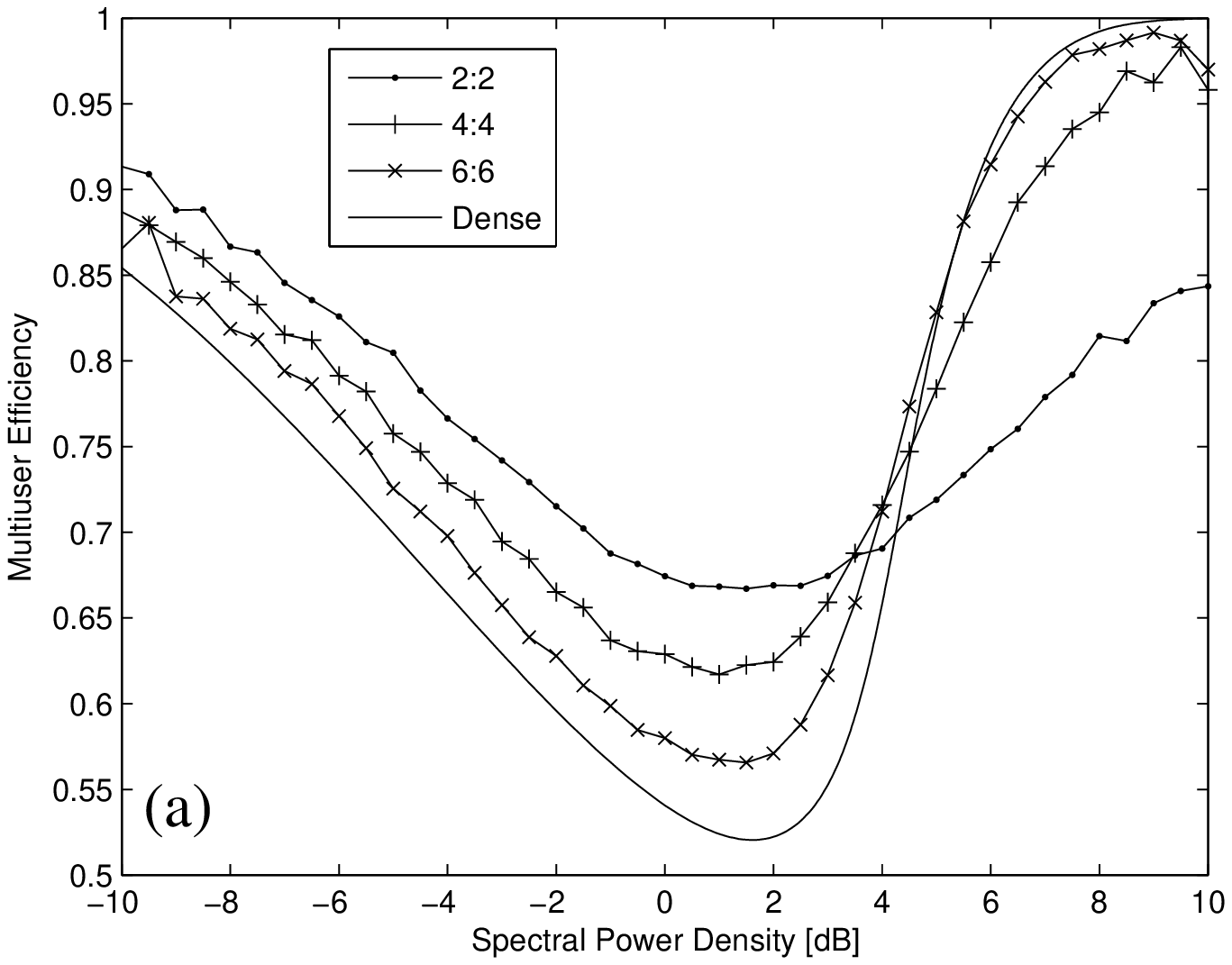}
    \epsfxsize=7.5cm
    \epsfbox{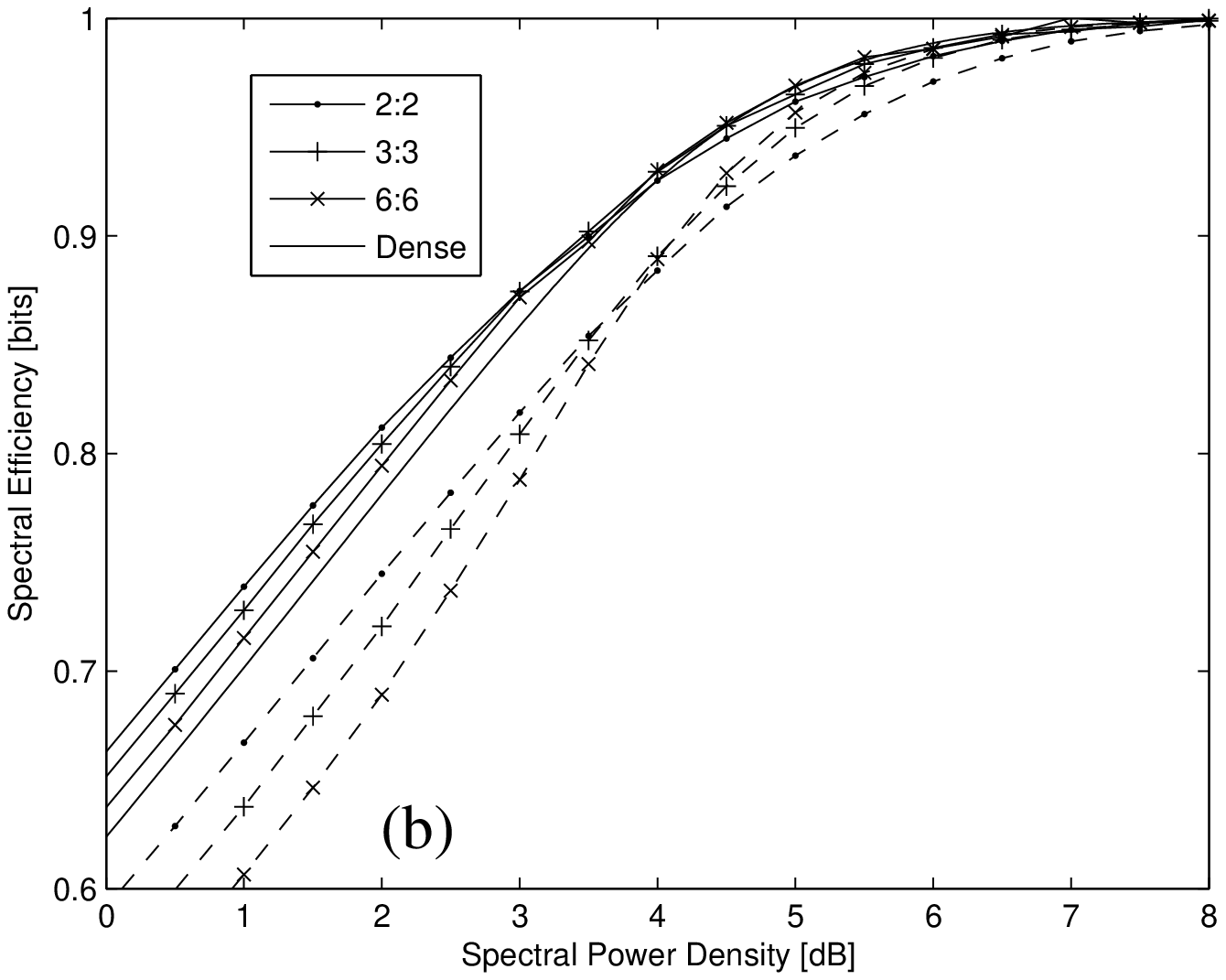}
  \end{center}
  \caption{\label{density_prop} The effect of increasing
density for the regular ensemble: (a) Multiuser efficiency,
    (b) spectral efficiency [$\full$] and mutual information [$\broken$].
Data presented on the basis of 10 runs, error bars are
    omitted but of a size comparable with the smoothness of the curves.
    The performance of sparse codes rapidly approaches that of the dense code everywhere.
    The $PSD$ threshold beyond which the dense code outperforms the sparse code is fairly stable.
  }
\end{figure}

The outperformance of dense codes by sparse ensembles with regular
user connectivity in the low $PSD$ regime is new to our knowledge, although Poissonian chip connectivity 
is everywhere inferior to both the dense and regular sparse codes. 
The difference between codes disappears rapidly with increasing (connection) density at fixed $\alpha$
(figure ~\ref{density_prop}). 
This is inline with our prediction of the regular code being a high performance ensemble in preceeding sections.

Figure~\ref{density_prop} indicates the effect of increasing
density at fixed $\alpha$ in the case of the regular code.
As density is increased the statistics of the sparse
codes approach that of the dense channel in all ensembles tested.
For the irregular ensemble performance increases
monotonically with density at all $PSD$. The rapid convergence to
the dense case performance was elsewhere observed for partly
regular ensembles, and ensembles based on a Gaussian prior
input~\cite{Yoshida:ASS,Montanari:ABP}. At all densities for 
which single solutions were found
the RS assumption appeared validated in the stability parameter and entropy.

Figure~\ref{alpha_prop} indicates the effect of channel load
$\alpha$ on performance. We first explain results for codes in
which only a single solution was found (no solution coexistence).
For small values of the load a monotonic increase in the bit error
rate, and capacity are observed as $\alpha$ is increased with $C$
constant, as shown in figures~\ref{alpha_prop}(a) and
~\ref{alpha_prop}(b), respectively. This matches the trend in the
dense case, the dense code becoming superior in performance to the
sparse codes as $PSD$ increases. We found that for all sparse ensembles
there existed regimes with $\alpha>1.49$ for which only a single
stable solution existed, although the equivalent dense systems are known to have
two stable solutions in some range of $PSD$~\cite{Tanaka:SMA}. In all
single valued regimes we observed positive entropy, and a negative
stability parameter. However, in cases of large $\alpha$ many
features became
more pronounced close to the dense case solution coexistence
regime: notably the cusp in the stability parameter, gap between
$MI$ and
$\nu$ and the gradient in $P_b$.

\subsection{Solution Coexistence Regimes}
\begin{figure}
  \begin{center}
    \leavevmode
    \epsfxsize=7.5cm
    \epsfbox{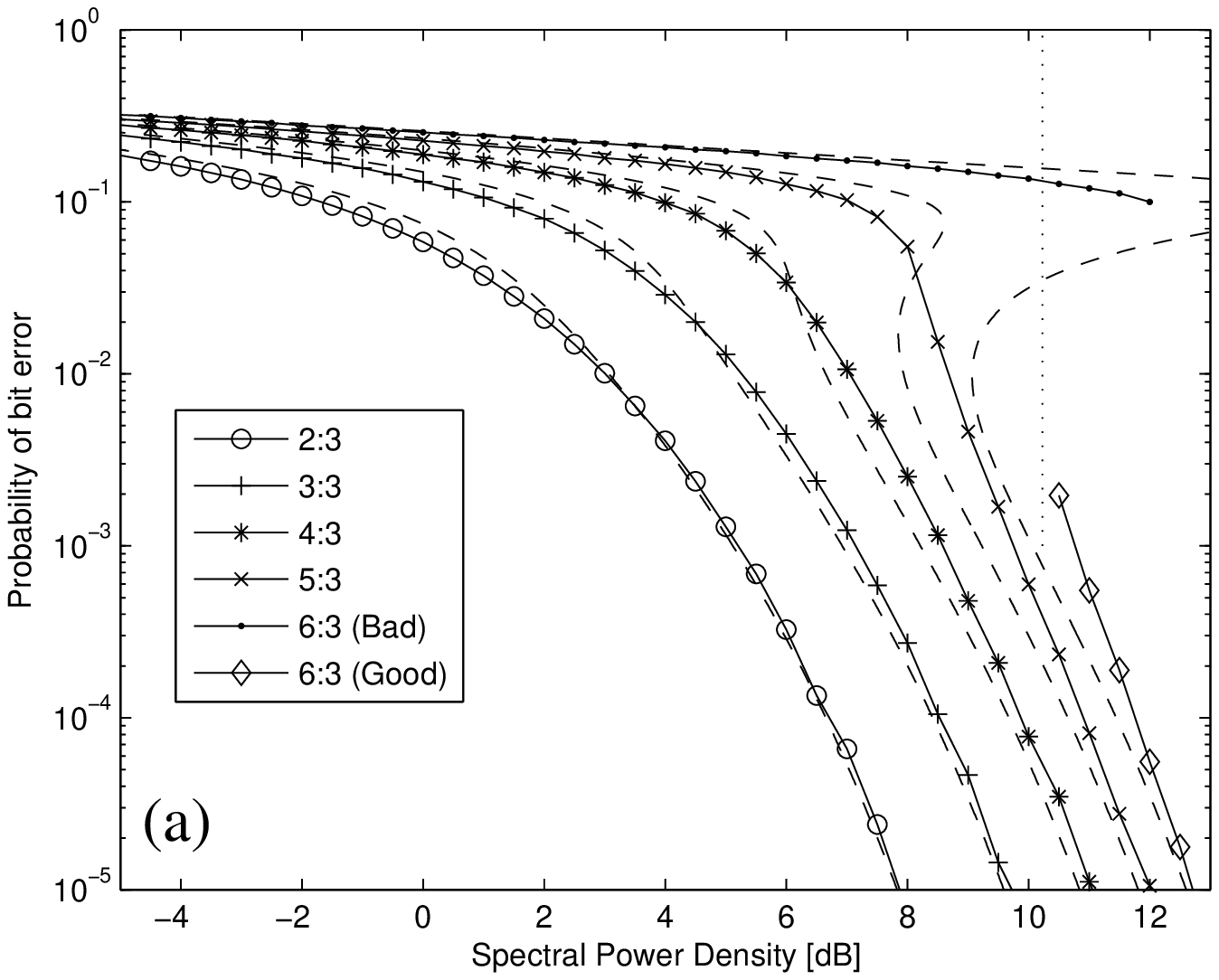}
   \epsfxsize=7.5cm
    \epsfbox{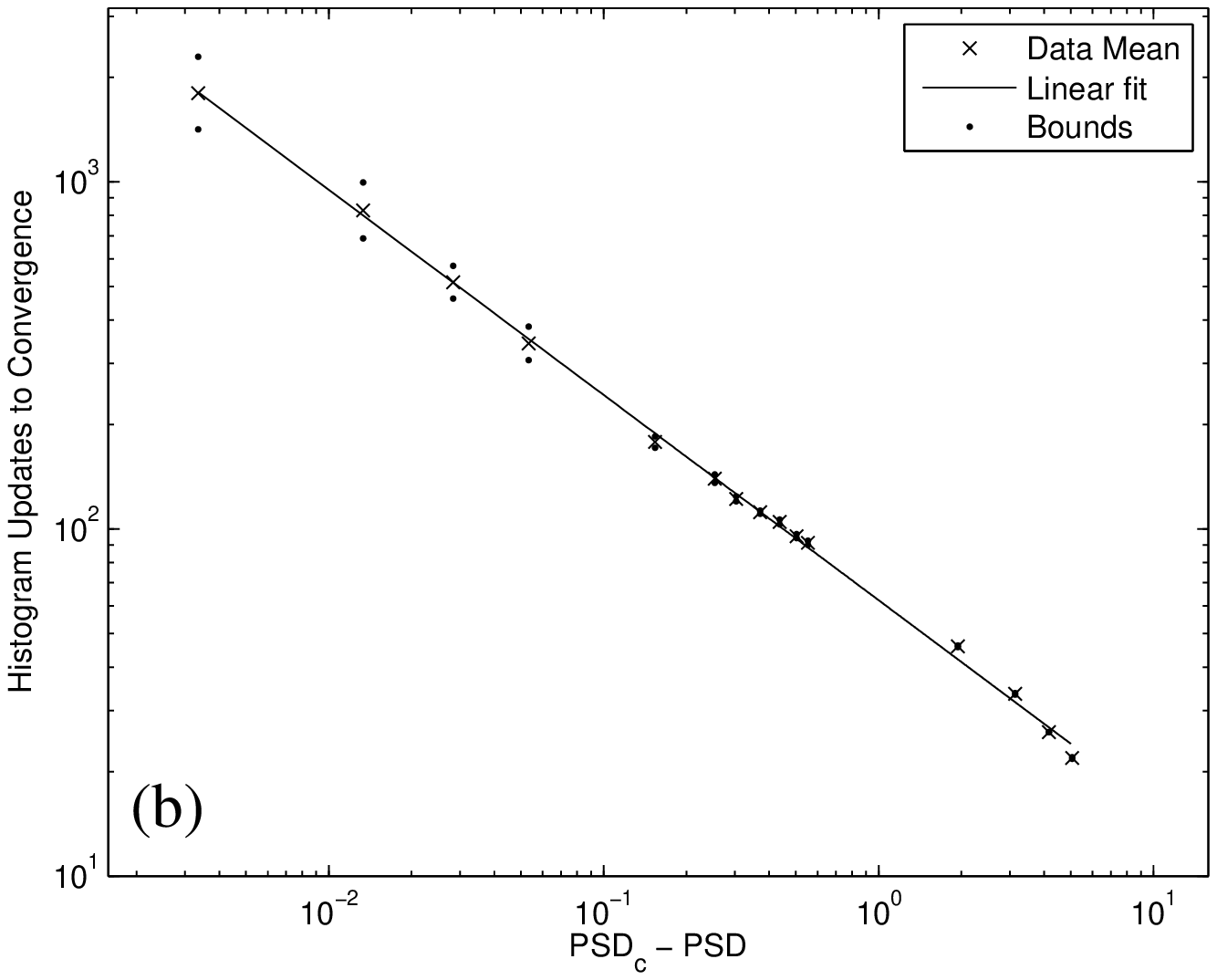}
   \epsfxsize=7.5cm
    \epsfbox{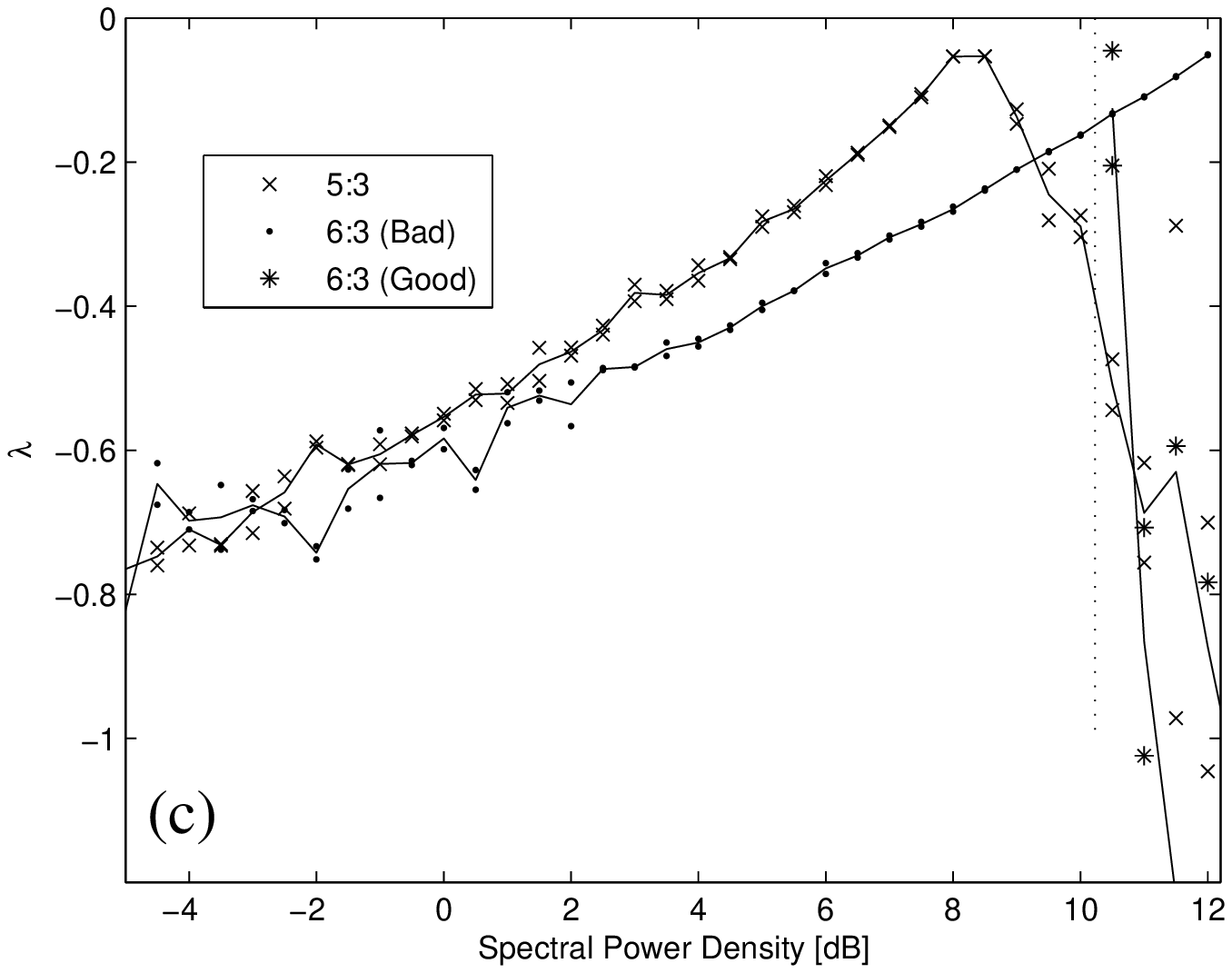}
    \epsfxsize=7.5cm
    \epsfbox{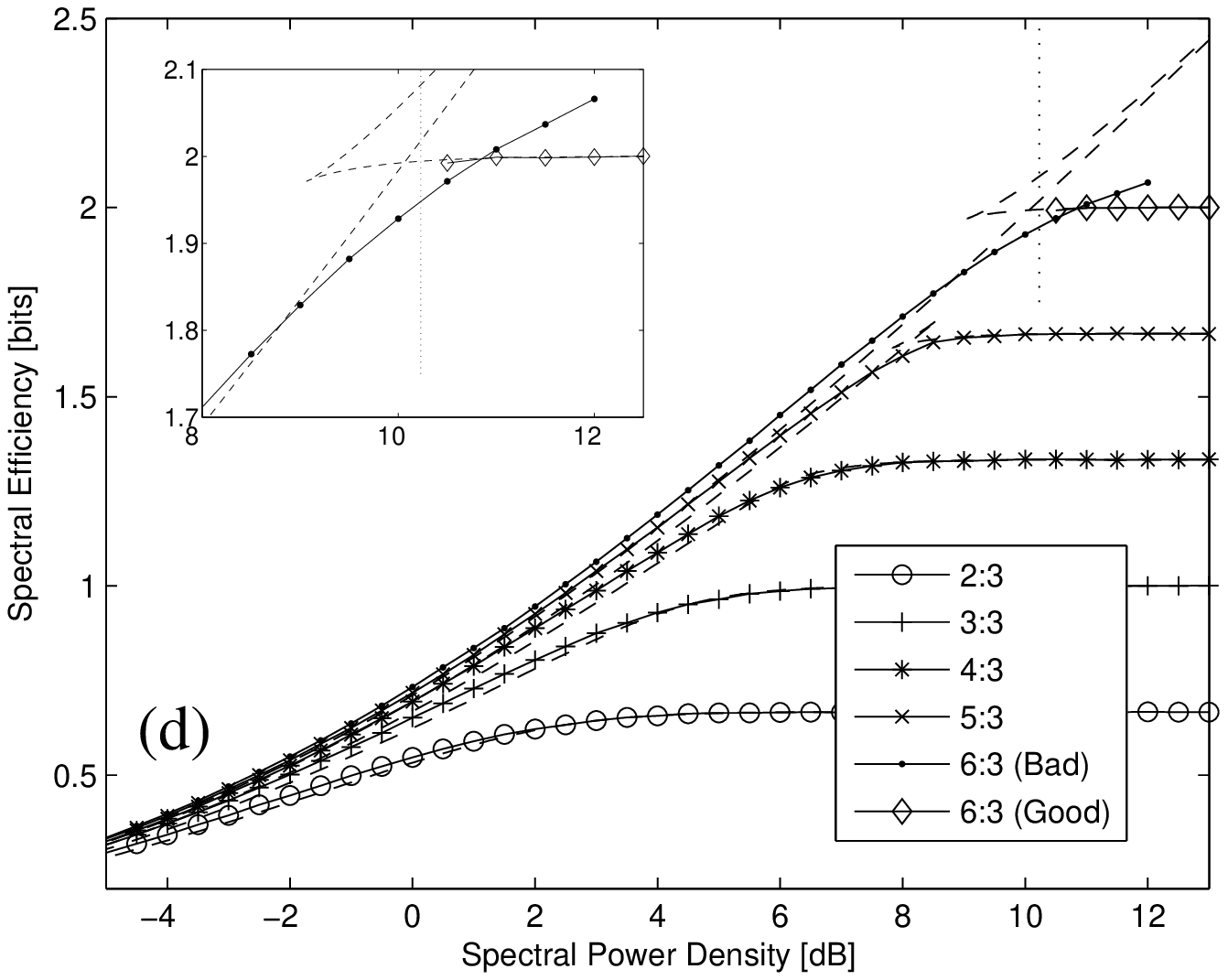}
  \end{center}
  \caption{\label{alpha_prop} The effect of channel load
  $\alpha$ on performance for the regular ensemble. Data presented on the basis of 10 runs,
  error bars omitted but characterised by the smoothness of curves. Dashed
  lines indicate the dense code analogues. The vertical dotted line
  indicates the point beyond which $6:3$ random and ferromagnetic initial conditions
  failed to converge to the same solution, both dynamically stable solutions are shown
  beyond this point.
  (a) There is a monotonic increase in bit error rate with the increasing load.
  (b) Investigation of the $6:3$ code ($\alpha=2$) indicates a divergence
  in convergence time as $PSD \rightarrow 10.23dB$ with
  exponent $0.59$ based on a simple linear regression of 15 points (each point
  is the mean of 10 independent runs). Beyond this point different initial
  conditions give rise to one of two solutions.
  (c) The stability parameter was found to be
  negative for all convergent solutions, indicating the suitability of RS.
  Where the solution is near ferromagnetic the stability measure becomes
  quickly very noisy (as shown for the $5:3$ and $6:3$ codes).
  (d) As load $\alpha$ is increased
  there is a monotonic increase in capacity. The spectral efficiency for
  the 'bad' solution exceeds 2 in a small interval (equivalent to negative
  entropy), similar to the behaviour reported for the dense case.
  }
\end{figure}

As in the case of dense CDMA~\cite{Tanaka:SMA}, also here we
observe a regime where two solutions, of quite different
performance, coexist. In order to investigate the regime where two
solutions coexist we investigated the states arrived at from
random and ferromagnetic initial conditions (giving bad and good solutions respectively). 
Separate heuristic
convergence criteria were found for the histograms, and these
seemed to work well for the good solution. For the bad solution 
we simply present results after
a fixed number of histogram updates ($500$) as all convergence
criteria tested appeared either too stringent, to require
experimentally inaccessible timescales, or did not capture the
asymptotic values for important quantities like entropy. We believe $500$ updates to be 
sufficiently conservative to capture the properties of these solutions however.

Figure~\ref{alpha_prop}(a) shows the dependence of the bit error
rate on the load, which is also equivalent to $L/C$. There is a
monotonic increase in bit error rate with the load and the
emergence and coexistence of two separate solutions above a
certain point; in the case of the $6:3$ code the point above which
the two solutions coexist is $PSD = 10.23dB$ as indicated by the
vertical dotted line.

We use the regular code $6:3$ to demonstrate the solution
coexistence found above some $PSD$ in various codes. 
The onset of the bimodal
distribution can be identified by the divergence in the
convergence time in the single solution regime (the time for the ferromagnetic and random
histograms to converge to a common distribution). The time for
this to occur, in a heuristically chosen statistic and accuracy,
is plotted in figure~\ref{alpha_prop}(b). By a naive linear
regression across 3 decades we found a power law exponent of
$0.59$ and a transition point of $PSD = 10.23dB$, but cannot
provide a goodness of fit measure
to this data. This would represent the point at which at least two
stable solutions co-exist. 

Beyond $PSD\approx12dB$ only one stable solution is found from
both random and ferromagnetic initial conditions, corresponding
statistically to a continuation of the good solution. A solution
which statistically resembles a continuation of the bad solution is
occasionally arrived at from both initial conditions, this solution
had a positive stability parameter and negative entropy -- so is
not a viable solution. Thus we predict a second
dynamical transition in the region of $12dB$, as might be guessed
by comparison with the dense case and observation of the trend in
the stability parameter (see figure~\ref{alpha_prop}(c)).

The stability results are presented in figure~\ref{alpha_prop}(c).
Only two stable solutions were found in the region beyond this
critical point and upto $12dB$, which we infer to be viable RS
solutions (where entropy is positive). The bad solution upto 12dB has a well resolved negative value. The good solution has a negative value in its mean, but like other near ferromagnetic solutions investigated results are very noisy due to numerical issues relating to histogram resolution.

Both capacity and spectral efficiency monotonically increase with
the load as shown in figure~\ref{alpha_prop}(d). For the $6:3$
code we see a separation of the two solutions at $PSD = 10.23dB$
(vertical dotted line.) The dashed lines correspond to a similar
behaviour observed in the dense case (the range of interest is
magnified in the inset.) A cross over in the entropy of the two
distinct solutions, near $PSD\approx11dB$, is indicative of a
second order phase transition. As in the dense case, only the solution of smallest spectral 
efficiency is thermodynamically relevant at a given $PSD$, although the other is
likely to be important in decoding dynamics. 
The trends in the sparse case follow the dense case qualitatively, with the good solution having performance only slightly worse than the corresponding solution in the dense case (and vice versa for the bad solution).

The entropy of the bad solution becomes negative in a small
interval (spectral efficiency exceeds 2) although no local instability
is observed. The static and dynamic properties of the histograms appear to be well resolved in this region.
However, the negative entropy indicates an instability towards either a type of solution not captured within the RS assumption, or towards some metastable configuration. We will not speculate further, the bad solution is in any case thermodynamically subdominant in its low and negative entropy form. 

Our hypothesis is therefore that the trends in the sparse
ensembles match those in the dense ensembles within the
coexistence region and $RS$ continues to be valid for each of two
distinct positive entropy solutions. The coexistence region for the sparse codes is
however smaller than in the corresponding dense ensembles. Since our histogram
updates mirror the properties of a belief propagation algorithm on
a random graph we can suspect that the bad solution may have implications for the
performance of belief propagation decoding in the
coexistence region, and that convergence problems will appear near
this region. In the user regular codes investigated the bad solution of the
sparse ensemble outperforms the bad solution of the dense
ensemble, and vice-versa for the good solution. Thus regardless of whether 
sparse decoding performance is good or bad, the dynamical transition point for the dense ensemble would
corresponds to a $PSD$ beyond which dense CDMA outperforms sparse
CDMA at a particular load.

\subsection{Spectral Efficiency Lower Bound Numerical Results}
\label{selbnr}
\begin{figure}
  \begin{center}
    \leavevmode
    \epsfxsize=7.5cm
    \epsfbox{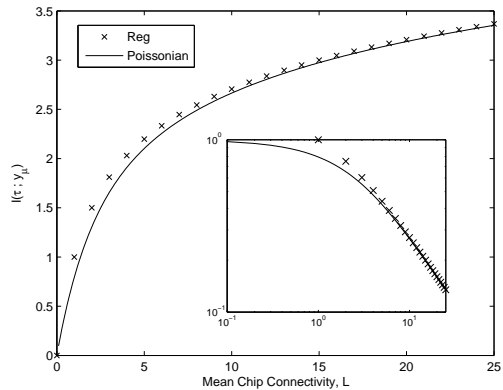}
  \end{center}
  \caption{\label{chipcapacity} A $PSD\rightarrow\infty$ limit to the expected mutual information between a single chip, and the transmitted bits. Mutual Information is highest for regular chip connectivities, with the Poissonian chip connectivity result also shown,
  the discrepancy becoming relatively small as $L$ increases. The inset shows the mutual information/bit decoded ($\left\langle I(\vtau;y_\mu)\right\rangle/L$) on a log-log plot to demonstrate an asymptotic power law behaviour and show more detail in the cases of small $L$.
  }
\end{figure}

Finally we present figure~\ref{chipcapacity}, which shows the
the mutual information between a single chip and transmitted bits for sparse ensembles of differing chip
connectivity in the infinite $PSD$ (zero noise) limit (\ref{entropylowerbound}). This shows
that in expectation a chip drawn from the regular ensemble contains more information on the transmitted bits than a chip drawn from any other ensemble (including the Poissonian ensemble).
The difference between the regular and Poissonian ensembles becomes relatively smaller as $L$ increases. 
This appears consistent with the replica method results found at high $PSD$, although regular chip
connectivity under performed by comparison with Poisson
distributed chip connectivity in the low $PSD$ regime, which was
not anticipated by the single chip approximation.

%%%%%%%%%%%%%%%%%%%%%%%%%%%%%%%%%%%%%%%%%%%%%%%%%%%%%%%
\section{Concluding Remarks}
\label{conclusion}
%%%%%%%%%%%%%%%%%%%%%%%%%%%%%%%%%%%%%%%%%%%%%%%%%%%%%%%
Our results demonstrate the feasibility of sparse regular codes
for use in CDMA. At moderate $PSD$ it seems the performance of
sparse regular codes may be very good. With the replica symmetric
assumption apparently valid at practical $PSD$ it is likely that
fast algorithms based on belief propagation may be very successful
in achieving the theoretical results. Furthermore for lower
density sparse codes the problem of the coexistence regime, which
limits the performance of practical decoding methods, seems to be
less pervasive than for dense ensembles in the over saturated
regime.

A direct evaluation of the properties of belief propagation may
prove similar results to those shown here. In the absence of
replica symmetry breaking states it is normally true that belief
propagation performs very well. However, to make best use of the
channel resources it may be preferable to implement high load
regimes in cases of high $PSD$, and so overcoming the algorithmic
problems arising from the solution coexistence is a challenge of
practical importance in this case.

Other practical issues in implementation are certainly
significant. Similar to the case of dense CDMA there are
considerable problems relating to multipath, fading and power
control, in fact it is known that these effects are more
disruptive for the sparse codes, especially regular codes.
However, certain situations such as broadcasting (one to many)
channels and downlink CDMA, where synchronisation can be assumed,
may be practical points for future implementation. There are
practical advantages of the sparse case over dense and orthogonal
codes in some regimes. The sparse CDMA method is likely to be
particularly useful in frequency-hopping and time-hopping code
division multiple access (FH and TH -CDMA) applications where the
effect of these practical limitations is less emphasised.

Extensions based on our method to cases without power control or
synchronisation have been attempted and are quite difficult. A
consideration of priors on the inputs, in particular the effects
when sparse CDMA is combined with some encoding method may also be
interesting.

%%%%%%%%%%%%%%%%%%%%%%%%%%%%%%%%%%%%%%%%%%%%%%%%%%%%%%%
\ack Support from EVERGROW, IP No.~1935 in FP6 of the EU is
gratefully acknowledged. DS would like to thank Ido Kanter for
helpful discussions.

\section*{Bibliography}
\bibliographystyle{unsrt}
\bibliography{Bibliography}

\end{document}